**The vacuum interpretation of quantum mechanics and the vacuum universe**

Ding-Yu Chung

In this paper, quantum mechanics is interpreted by the adjacent vacuum that behaves as a virtual particle.  Matter absorbs and emits its adjacent vacuum, constituting the formations of particle and wave. As described in the vacuum universe model, the adjacent vacuum is derived from the pre-inflationary universe in which the pre-adjacent vacuum is absorbed by the pre-matter. This absorbed pre-adjacent vacuum is emitted to become the added space for the inflation in the inflationary universe whose space-time is separated from the pre-inflationary universe.  This added space is the adjacent vacuum. The absorption of the adjacent vacuum as the added space results in the adjacent zero space (no space), Quantum mechanics is the interaction between matter and the three different types of vacuum: the adjacent vacuum, the adjacent zero space, and the empty space. The absorption of the adjacent vacuum results in the empty space superimposed with the adjacent zero space, confining the matter in the form of particle. When the absorbed vacuum is emitted, the adjacent vacuum can be anywhere instantly in the empty space superimposed with the adjacent zero space where any point can be the starting point (zero point) of space-time. Consequently, the matter that expands into the adjacent vacuum has the probability to be anywhere instantly in the form of wavefunction.  In the vacuum universe model, the universe not only gains its existence from the vacuum but also fattens itself with the vacuum.  During the inflation, the adjacent vacuum as the added space also generates the periodic table of elementary particles to account for all elementary particles and their masses in a good agreement with the observed values.  The universe ends with the return of the adjacent vacuum to the pre-inflationary universe, and it is subsequently absorbed to start another new inflationary universe. The universe is an endless fattening free lunch.

## 1.    *Introduction*

There are many interpretations of quantum mechanics.  In this paper, quantum mechanics is explained by the adjacent vacuum that behaves as a virtual particle. Matter absorbs and emits its adjacent vacuum, constituting the formations of particle and wave.  As described in the vacuum universe model [1], the adjacent vacuum is derived from the pre-inflationary universe in which the pre-adjacent vacuum is absorbed by the pre-matter. This absorbed pre-adjacent vacuum is emitted to become the added space for the inflation in the inflationary universe whose space-time is separated from the pre-inflationary universe. In this inflationary universe, this added space is the adjacent vacuum. The absorption of the adjacent vacuum as the added space results in the adjacent zero space (no space), Quantum mechanics is the interaction between matter and the three different types of vacuum: the adjacent vacuum, the adjacent zero space, and the empty space. The absorption of the adjacent vacuum results in the empty space superimposed with the adjacent zero space, confining the matter in the form of particle. When the absorbed vacuum is emitted, it can be anywhere instantly in the empty space superimposed with the adjacent zero space where any point can be the starting point (zero point) of space-time.  Consequently, the matter that expands into the adjacent vacuum has the probability to be anywhere instantly in the form of wavefunction.

In the vacuum universe model, the universe not only gains its existence from the vacuum but also fattens itself with the vacuum.  During the inflation, the absorbed pre-adjacent vacuum is emitted to become instantly the add space for the inflation. In terms of elementary particles, the added space for the inflation is used to dilute (fractionalize) the high mass mixed 9-branes.  The diluted products are the hierarchical mixed branes from 9 to 3 whose masses decrease with the space-time dimension numbers as in the dimensional hierarchy [2].

In the dimensional hierarchy, the masses of space dimensions follow the hierarchical dimensional mass formula based on the Planck mass and the space-time dimension numbers as follows.

$$M_D = M_P \, \alpha^{2(11-D)} \qquad (1)$$

where $M_D$ is mass of a dimension, D is space-time dimension number, $M_P$ is the Planck mass, and $\alpha$ is fine structure constant, the probability of a fermion emitting or absorbing a boson. The mass of the eleventh dimension is the Planck mass.

The hierarchical dimensional mass formula is derived from the assumption that each space dimension is occupied by a dimensional fermion $F_D$ and a dimensional boson $B_D$.  The probability to transforming a fermion into its boson in the adjacent dimension is same as the fine structure constant, $\alpha$, the probability of a fermion emitting or absorbing a boson.  The probability to transforming a boson into its fermion partner in the same dimension is also the fine structure constant, $\alpha$.  This hierarchy is expressed in term of the dimension space-time number, D,

$$M_{D-1, B} = M_{D, F} \, \alpha_{D, F}, \qquad (2)$$

$$M_{D, F} = M_{D, B} \, \alpha_{D, B}, \qquad (3)$$

where $M_{D, B}$ and $M_{D, F}$ are the masses for a dimensional boson and a dimensional fermion, respectively, and $\alpha_{D, B}$ or $\alpha_{D, F}$ is the fine structure constant, which is the ratio between the energies of a dimensional boson and its dimensional fermionic partner. Assuming $\alpha$ is the same for all dimensional fermions and dimensional bosons, Eq. (1) is derived.  (As shown later, with one exception, all $\alpha$'s are equal to $\alpha_e$, the fine structure constant for the electromagnetic field, so Eq. (1) is only approximately correct.)

The process of interbrane-vacuum mixing is described in Section 2 for the pre-inflationary universe and the inflationary universe.  After the inflation, the observable universe consists of the adjacent vacuum, the zero space, the empty space, the hierarchical mixed branes from the mixed 3- to 9- branes, and force fields.  The ordinary universe consists of the mixed 3-brane as the mixture of leptons and quarks. In Section 3, the periodic table of elementary particles is constructed to account of all leptons, quarks, gauge bosons, hadrons, and their masses in a good agreement with the observed values. In Section 4, the cyclic universe is proposed.  The conclusion is that quantum mechanics and the periodic table of elementary particles are derived from the vacuum universe model.

## 2. *The Interbrane-vacuum Mixing: the Pre-inflation and the Inflation*

The pre-inflationary universe is derived from a vacuum fluctuation.  In this paper, the pre-inflationary universe consists of an eleven space-time with two identical boundary 9-branes separated by a finite gap spanning a bulk space for gravity.  The bulk space is the pre-adjacent vacuum, and the boundary 9-brane is the pre-matter.  It is based on the Horava-Witten model on eleven dimensional supergravity on a manifold with boundary [3].  Basically, it is a joining of two eleven dimensional membranes.   In the ekpyrotic universe model [4] and its modification, the pyrotechnic universe model [5] based on Horava-Witten theory, the boundary 3-branes (hidden and visible 3-branes) are different.  There are a number of other models [6] for brane inflation.

The inflationary universe starts with the interbrane-vacuum mixing between the boundary 9-branes and between the 9-branes and the pre-adjacent vacuum (bulk space).  As shown later, this mixing is the most fundamental mixing, combining the space dimensions from the two boundary 9-branes and combining the 9-branes and the pre-adjacent vacuum (bulk space).  It is analogous to the combining of two n units DNA to form 2n units of DNA during sexual reproduction to generate a new life.  The product is the mixed 9-brane that has the combined space dimensions from both boundary 9-branes and the absorbed pre-adjacent vacuum.  The inflationary universe has a new set of space-time, separated from the space-time of the pre-inflationary universe.

Symmetrical to the pre-inflationary universe, the inflationary universe consists of the two boundary mixed 9-branes with gravity during the interbrane-vacuum mixing as in Fig. 1. (This inter-universal symmetry later becomes the pattern for the symmetry between the electric and the magnetic fields.)

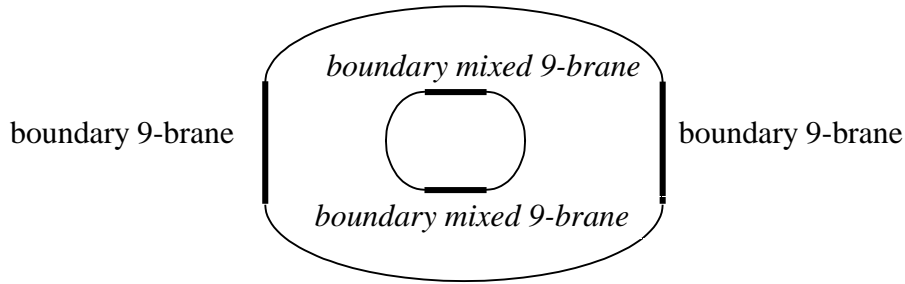

**Fig. 1:** the inter-universal symmetry between the pre-inflationary universe with the two boundary 9-branes and the inflationary universe with the two boundary mixed 9-branes

This inter-universal symmetry prevents the absorbed pre-adjacent vacuum in the mixed 9-branes to convert the boundary mixed 9-branes into the low mass mixed branes by the vacuum dilution (fractionalization). The symmetry behaves as a high pressure to keep bubbles (the absorbed pre-adjacent vacuum) from expanding.

As soon as all of the boundary 9-branes are depleted, matter emits the absorbed pre-adjacent vacuum. This absorbed pre-adjacent vacuum is emitted to become the added space for the inflation in the inflationary universe. In the inflationary universe, this added space is the adjacent vacuum.

During the inflation, this added space for the inflation is used to dilute (franctionalize) the high mass boundary mixed 9-branes into the low mass hierarchical mixed branes. This vacuum dilution process has two different modes: the big band mode for the hidden universe and the big bang mode for the observable universe as Fig. 2 and Fig.3.

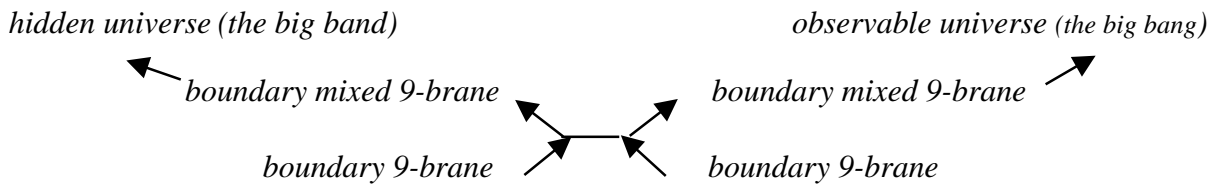

**Fig. 2:** the historical diagram for the formation of the observable and hidden universes

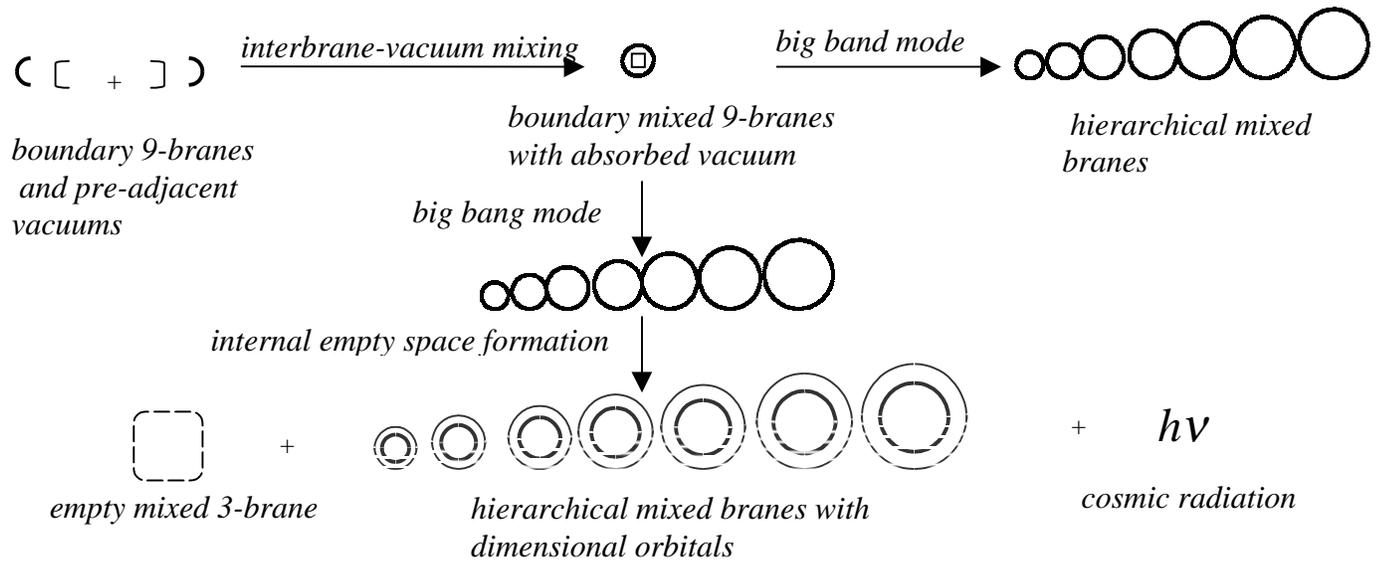

**Fig. 3:** the interbrane-vacuum mixing and the vacuum dilution process (the big band mode and the big bang mode)

In the big band mode, the mixed 9-brane is converted into the hierarchical mixed branes from 3 to 9, whose masses follow the hierarchical dimensional mass formula as Eq. (1) where p-brane has p + 1 space-time dimension, D. Each mixed brane has an attached massive gravity. All mixed branes are adjacent to one another. There is no empty space internally among the mixed branes in the big band mode.

In the big bang mode, the formation of the hierarchical mixed branes is followed by the internal empty space formation, which provides gaps among the mixed branes. The mechanism to generate the empty space internally is to assign symmetrical and opposite charges to the two internal boundary branes within the mixed brane, the resulting internal annihilation (implosion) dislocate energy from the mixed brane, and the internal empty space (the empty mixed 3-brane) is formed. The dislocated energy is cosmic radiation resided in the empty mixed 3-brane (the empty space). The attached gravity to the mixed branes that are annihilated changes from the massive form to the massless form of negative energy for the corresponding released positive energy from the mixed branes.

The mixed branes that are not annihilated have asymmetrical charges (CP nonconservation), in such way that the mixed brane has two asymmetrical sets (main and auxiliary) of space dimensions from the two boundary branes within the mixed brane. The auxiliary set is dependent on the main set, so the mixed brane appears to have only one set of space dimensions. The attached gravity to the mixed branes that are not annihilated is also in the massless form of negative energy for the corresponding positive brane mass.

Quantum mechanics is the interaction between matter and the three different types of vacuum: the adjacent vacuum, the adjacent zero space, and the empty space. When the adjacent vacuum as the added space is absorbed by a matter, the internal empty space is superimposed with the adjacent zero space (no space), confining the matter in the form of particle. When the absorbed vacuum is emitted, it can be anywhere instantly in the empty space superimposed with the adjacent zero space where any point can be the starting point (zero point) of space-time. Consequently, the matter that expands into the adjacent vacuum has the probability to be anywhere instantly in the form of wavefunction. Each wavefunction is specific to a particular adjacent vacuum, so the entanglement of different matters leads to the instant collapse of this wavefunction and the instant adsorption of the adjacent vacuum to form an entangled matter with the entangled adsorbed vacuum, which can then be emitted to bring about a new wavefunction.

The internal empty space in the big bang mode allows all the mixed branes to have the same number of space dimensions. It is achieved by adding the virtual space dimensions to the empty space surrounding the core mixed branes in the manner of the Kaluza-Klein structure, where the virtual space dimension is a one-dimensional circle associated with every point in the core brane dimension, and every higher virtual dimension circles the circle of the lower virtual dimension. The virtual space dimensions are the space dimensions in between the core brane space dimension and the gravity (the eleventh space-time dimension). The masses of the space dimensions follow the hierarchical dimensional mass formula as Eq. (1).

The core mixed brane absorbs the hierarchical virtual dimensions as the scalar fields through Higgs mechanism. The surrounding hierarchical virtual dimensions are converted into the "dimensional orbitals" with the same space-time dimension as the core mixed brane. It is analogous to the conversion of the material from a collapsed tall three-dimensional building into the material to extend the land of a virtually two-dimensional flat seashore.

When there is not enough mass-energy in the core brane for the masses of the dimensional orbitals, there are continuous emission and absorption of virtual particles from and to the core mixed branes. Gravity resided in the empty space is also a part of the hierarchical dimensional orbitals for the mixed branes. These dimensional orbitals are the force fields for the mixed branes in the big bang mode.

Since the core mixed branes have two sets (main and auxiliary) of space dimensions, there are also two sets of the dimensional orbitals. For the mixed 3-brane in the big bang mode, there are two sets of the hierarchical seven dimensional orbitals (including gravity) and the non-hierarchical three core brane space dimensions. The mixed 3-brane is the mixture of leptons and quarks.

The big band mode is used in the hidden universe. There is no internal empty space in the big band mode, so the 9-mixed branes is fractionalized into lower mixed branes slowly and sequentially to avoid any rupture internally to create the internal empty space. It is manifested in the extremely slow stepwise fractionalization and condensation from the mixed 9-brane to the mixed 3-brane and back to the mixed 9-brane, resulting in expansion and contraction, like a big elastic rubber band (big band).

The big bang mode is used in the observable universe. The cosmic way to obtain the internal empty space in the big bang mode is to stretch the universe rapidly in a "superluminal expansion" to achieve a "rupture", resulting in the formation of an "internal gap" as the internal empty space. Therefore, the empty space and cosmic radiation emerge only after the superluminal inflationary emergence of all of the hierarchical mixed branes. Consequently, the inflationary emergence of the mixed branes and the non-inflationary emergence of cosmic radiation constitute the hybrid inflation [7].

The two universes have parallel sets of space-time. The gravity of the hidden universe other than the gravity with the mixed 3-brane is hidden from the observable universe because it is massive, and is not compatible.

### 3. *The Ordinary Universe: the Periodic Table of Elementary Particles*

The observable universe consists of the mixed branes from 3 to 9. The ordinary (baryonic) universe in the observable universe consists of the mixed 3-brane, which is the mixture of leptons and quarks. Exotic dark matter in the observable universe consists of the mixed branes from 4 to 9. As shown later, exotic dark matter cannot be seen, but it can be observed by gravity. The ordinary (baryonic) matter is one of the seven mixed branes at equal mass proportions, so the baryonic mass fraction is 1/7 (0.14). The universal baryonic mass fraction was found to be 0.13 by the observations of primordial deuterium abundance [8]. The calculated value agrees well with the observed value.

For ordinary matter (the mixed 3-brane), there are two sets (main and auxiliary) of the seven dimensional orbitals. The total number of dimensional orbitals is 14 as shown in Fig. 4.

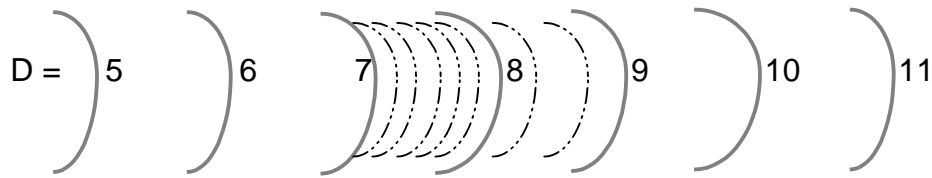

**Fig. 4:** 14 dimensional orbitals in the mixed 3-brane: 7 main dimensional orbitals (solid line), and 7 auxiliary dimensional orbitals (dot line), D = main dimensional orbital number

As shown in Fig. 4, the fifth main dimensional orbital is the start of the main dimensional orbitals. To be adjacent to the start of the auxiliary dimensional orbitals, the seventh main dimensional orbitals mixes with the fifth orbitals. Such mixing is manifested by the symmetry mixing in the electroweak interaction between U (1) for the fifth main dimensional orbital and SU (2) for the seventh main dimensional orbitals.

Other than gravity (the eleventh main dimensional orbital that relates to mass and energy only), each dimensional orbital carries certain discrete functions as the manifestation of cosmology. The main dimensional orbitals are for all major functions, and the auxiliary dimensional orbitals are for mostly quarks. Each main dimensional orbital except gravity is assigned to carry gauge symmetry and space-time symmetry.

The fifth main dimensional orbital with U(1) gauge symmetry carries the charge for the two symmetrical internal boundary branes within a mixed brane. During the annihilation in the big bang mode of the vacuum dilution process, the two symmetrical internal boundary branes become massless energy, so the fifth main dimensional orbital is a massless orbital. All other dimensional orbitals except gravity are short-range massive dimensional orbitals. Only the mixed 3-brane (ordinary matter) has the fifth main dimensional orbital, so without the fifth main dimensional orbital, exotic dark matter consists of permanently neutral higher dimensional particles. It cannot emit light, and cannot form atoms.

The sixth main dimensional orbital carries the charge for the asymmetrical internal boundary branes within a mixed brane. The dependence of the auxiliary dimensional orbitals (quarks) on the main dimensional orbitals (leptons) is represented by the colors as in color SU(3) to become colorless as in U(1), so a quark composite must have integer charge and hypercharge, like leptons, in order to have an independent existence.

The seventh main dimensional orbital carries the charge for the gauge symmetry within the family of leptons or quarks. The charge is represented by SU(2) for the symmetry between two leptons (e and ν) or between two quarks (u and d). The same set of gauge symmetry groups is assigned to the eighth, the ninth, and the tenth main dimensional orbitals as U(1), U(1), and SU(2), respectively.

Various space-time symmetries reflect the existences of fermions. P nonconservation is required to achieve chiral symmetry for massless leptons (neutrinos), so masses of higher neutrinos in the lepton family are not too large to fit in the lepton family. In order to have a long-term existence of fermions, CP nonconservation is required to distinguish matter from anti-matter. P and CP nonconservations are in pairs of the right and the left. The seventh, the eighth, the ninth, and the tenth main dimensional orbitals are assigned to have P (left), CP (right), CP (left), and P (right) nonconservation, respectively. There is no nonconservation for the fifth and the sixth main dimensional orbitals.

The combination of the gauge symmetry and the space-time symmetry for the main dimensional orbitals from 5 to 10 results in U(1), SU(3) to become U(1), SU(2)$_L$, U(1)$_R$, U(1)$_L$, and SU(2)$_R$, respectively. The eleventh main dimensional orbital is for gravity in the massless form as the negative energy for the corresponding positive energy and mass.

The lack of perfect symmetry between electric and magnetic fields is a reflection of the lack of perfect symmetry between the boundary 9-branes and the boundary mixed 9-branes as in Fig. 1. The two boundary 9-branes represent two electric charges. The mixing (moving) of the boundary 9-branes generates the two boundary mixed 9-branes, representing two magnetic charges. The observable universe inherits only one boundary mixed 9-brane, which has to assume the role of two magnetic charges at the same time. Therefore, there are separable electric charges, but no separable magnetic charges.

The structure of the mixed 3-brane with dimensional orbitals resembles to the structure of atomic orbital. Consequently, the periodic table of elementary particles is constructed to account of all leptons, quarks, gauge bosons, and hadrons as described in details in Reference 2. It is briefly reviewed here.

For the gauge bosons, the seven main dimensional orbitals are arranged as $F_5 B_5 F_6 B_6 F_7 B_7 F_8 B_8 F_9 B_9 F_{10} B_{10} F_{11} B_{11}$, where B and F are boson and fermion in each orbital. The masses of the main dimensional orbital bosons can be derived from Eqs. (2) and (3). Assuming $\alpha_{D,B} = \alpha_{D,F}$, the relation between the bosons in the adjacent main dimensional orbitals, then, can be expressed in terms of the main dimensional orbital number, D,

$$M_{D-1, B} = M_{D, B} \, \alpha^2_D , \qquad (4)$$

where D= 6 to 11, and $E_{5,B}$ and $E_{11,B}$ are the energies for the main dimensional orbital five and the main dimensional orbital eleven, respectively. The lowest energy is the Coulombic field, $E_{5,B} = \alpha M_{6,F} = \alpha M_e$,

The bosons generated are the main dimensional orbital bosons or $B_D$. Using only $\alpha_e$, the mass of electron, the mass of $Z^0$, and the number (seven) of dimensional orbitals, the masses of $B_D$ as the gauge boson can be calculated as shown in Table 1.

**Table 1.** The Masses of the main dimensional orbital bosons:
$\alpha = \alpha_e$, D = main dimensional orbital number

| $B_D$ | $M_D$ | GeV (calculated) | gauge boson | Interaction, symmetry |
|---|---|---|---|---|
| $B_5$ | $M_e \, \alpha$ | $3.7 \times 10^{-6}$ (given) | A | electromagnetic, U(1) |
| $B_6$ | $M_e/\alpha$ | $7 \times 10^{-2}$ | $\pi_{1/2}$ | strong, SU(3) $\longrightarrow$ U(1) |
| $B_7$ | $M_6/\alpha_w^2 \cos\theta_w$ | 91.177 (given) | $Z_L^0$ | weak (left), SU(2)$_L$ |
| $B_8$ | $M_7/\alpha^2$ | $1.7 \times 10^6$ | $X_R$ | CP (right) nonconservation, U(1)$_R$ |
| $B_9$ | $M_8/\alpha^2$ | $3.2 \times 10^{10}$ | $X_L$ | CP (left) nonconservation, U(1)$_L$ |
| $B_{10}$ | $M_9/\alpha^2$ | $6.0 \times 10^{14}$ | $Z_R^0$ | weak (right), SU(2)$_R$ |
| $B_{11}$ | $M_{10}/\alpha^2$ | $1.1 \times 10^{19}$ | G | gravity |

In Table 1, $\alpha = \alpha_e$ (the fine structure constant for electromagnetic field), and $\alpha_w$ is not same as $\alpha$ of the rest, because there is a mixing between $B_5$ and $B_7$ as the symmetry mixing between U(1) and SU(2) in the standard theory of the electroweak interaction, and $\sin\theta_w$ is not equal to 1. As shown in Reference 2, $B_5$, $B_6$, $B_7$, $B_8$, $B_9$, and $B_{10}$ are A (massless photon), $\pi_{1/2}$, $Z_L^0$, $X_R$, $X_L$, and $Z_R^0$, respectively, responsible for the electromagnetic field, the strong interaction, the weak (left handed) interaction, the CP (right handed) nonconservation, the CP (left handed) nonconservation, and the P (right handed) nonconservation, respectively. The calculated value for $\theta_w$ is $29.69^0$ in good agreement with $28.7^0$ for the observed value of $\theta_w$ [9]. The calculated energy for $B_{11}$ is $1.1 \times 10^{19}$ GeV in good agreement with the Planck mass, $1.2 \times 10^{19}$ GeV.

The two sets of seven dimensional orbitals result in 14 dimensional orbitals (Fig. 5) for gauge bosons, leptons, and quarks. The periodic table for elementary particles is shown in Table 2.

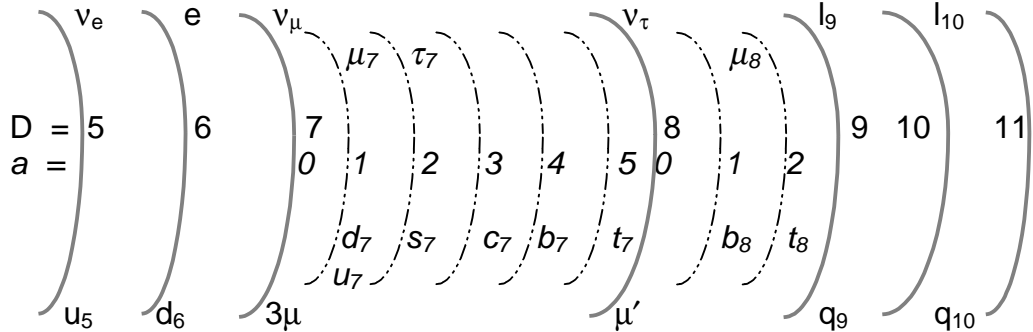

**Fig. 5.** leptons and quarks in the dimensional orbits
D = main dimensional number, a = auxiliary dimensional number

**Table 2.** The Periodic Table of elementary particles
D = main dimensional orbital number, a = auxiliary dimensional orbital number

| D | a = 0 | 1 | 2 | a = 0 | 1 | 2 | 3 | 4 | 5 | |
|---|---|---|---|---|---|---|---|---|---|---|
|   | Lepton |   |   | Quark |   |   |   |   |   | Boson |
| 5 | $l_5 = \nu_e$ |   |   | $q_5 = u_5 = 3\nu_e$ |   |   |   |   |   | $B_5 = A$ |
| 6 | $l_6 = e$ |   |   | $q_6 = d_6 = 3e$ |   |   |   |   |   | $B_6 = \pi_{1/2}$ |
| 7 | $l_7 = \nu_\mu$ | $\mu_7$ | $\tau_7$ | $q_7 = 3\mu$ | $u_7/d_7$ $s_7$ |   | $c_7$ | $b_7$ | $t_7$ | $B_7 = Z_L^0$ |
| 8 | $l_8 = \nu_\tau$ | $\mu_8$ |   | $q_8 = \mu'$ | $b_8$ | $t_8$ |   |   |   | $B_8 = X_R$ |
| 9 | $l_9$ |   |   | $q_9$ |   |   |   |   |   | $B_9 = X_L$ |
| 10 | $F_{10}$ |   |   |   |   |   |   |   |   | $B_{10} = Z_R^0$ |
| 11 | $F_{11}$ |   |   |   |   |   |   |   |   | $B_{11} = G$ |

D is the dimensional orbital number for the seven main dimensional orbitals. The auxiliary dimensional orbital number, a, is for the seven auxiliary dimensional orbitals, mostly for subquarks. All gauge bosons, leptons, and subquarks are located on the seven dimensional orbitals and seven auxiliary dimensional orbitals. Quarks and heavy leptons ($\mu$ and $\tau$) are in seven auxiliary spatial dimensions. Most leptons are dimensional orbital fermions, while all quarks are the sums of subquarks.

The fermion mass formula for massive leptons and quarks is derived from Reference 2 as follows.

$$M_{F_{D,a}} = \sum M_{F_{D,0}} + M_{AF_{D,a}}$$

$$= \sum M_{F_{D,0}} + \frac{3}{2} M_{B_{D-1,0}} \sum_{a=0}^{a} a^4 \quad (5)$$

$$= \sum M_{F_{D,0}} + \frac{3}{2} M_{F_{D,0}} \alpha_D \sum_{a=0}^{a} a^4$$

Each fermion can be defined by dimensional orbital numbers (D's) and auxiliary dimensional orbital numbers (a's). The compositions and calculated masses of leptons and quarks are listed in Table 3.

**Table 3.** The Compositions and the Constituent Masses of Leptons and Quarks
D = main dimensional orbital number and a = auxiliary dimensional orbital number

|  | $D_a$ | Composition | Calc. Mass |
|---|---|---|---|
| Leptons | $D_a$ for leptons |  |  |
| $\nu_e$ | $5_0$ | $\nu_e$ | 0 |
| e | $6_0$ | e | 0.51 MeV (given) |
| $\nu_\mu$ | $7_0$ | $\nu_\mu$ | 0 |
| $\nu_\tau$ | $8_0$ | $\nu_\tau$ | 0 |
| $\mu$ | $6_0 + 7_0 + 7_1$ | $e + \nu_\mu + \mu_7$ | 105.6 MeV |
| $\tau$ | $6_0 + 7_0 + 7_2$ | $e + \nu_\mu + \tau_7$ | 1786 MeV |
| Quarks | $D_a$ for quarks |  |  |
| u | $5_0 + 7_0 + 7_1$ | $u_5 + q_7 + u_7$ | 330.8 MeV |
| d | $6_0 + 7_0 + 7_1$ | $d_6 + q_7 + d_7$ | 332.3 MeV |
| s | $6_0 + 7_0 + 7_2$ | $d_6 + q_7 + s_7$ | 558 MeV |
| c | $5_0 + 7_0 + 7_3$ | $u_5 + q_7 + c_7$ | 1701 MeV |
| b | $6_0 + 7_0 + 7_4$ | $d_6 + q_7 + b_7$ | 5318 MeV |
| t | $5_0 + 7_0 + 7_5 + 8_0 + 8_2$ | $u_5 + q_7 + t_7 + q_8 + t_8$ | 176.5 GeV |

There are only three generations of leptons and quarks, because according to calculation, only three generations of quarks can fit in exactly seven auxiliary dimensional orbitals. The calculated masses are in good agreement with the observed constituent masses of leptons and quarks [10,11]. The mass of the top quark found by Collider Detector Facility is 176 ± 13 GeV [10] in a good agreement with the calculated value, 176.5 GeV.

As shown in Reference 2, the masses of hadrons can also be calculated based on binding energy derived from the auxiliary dimensional orbitals. The calculated values for the masses of hadrons are in good agreement with the observed values.

The masses of leptons, quarks, gauge bosons, and hadrons are calculated with only four known constants: the number of spatial dimensions, the mass of electron, the mass of Z°, and $\alpha_e$. Most importantly, the calculation shows that exactly seven main and seven auxiliary dimensional orbitals are needed for all fundamental interactions, leptons, quarks, and hadrons.

## 4. *The Cyclic Universe*

The hidden universe and the observable universe have parallel sets of space-time. The hidden universe is hidden from the observable universe until the quintessence transition [12], when the hidden universe fractionalizes into mixed 3-brane, compatible with the empty mixed 3-brane in the observable universe. The empty mixed 3-brane provides the place for the gravity in the hidden universe to be a massless force field as the gravity in the observable universe. It is manifested as the fifth dimension bulk space for massless anti-gravity as in the Randall - Sundrum mechanism [13] as Fig.6.

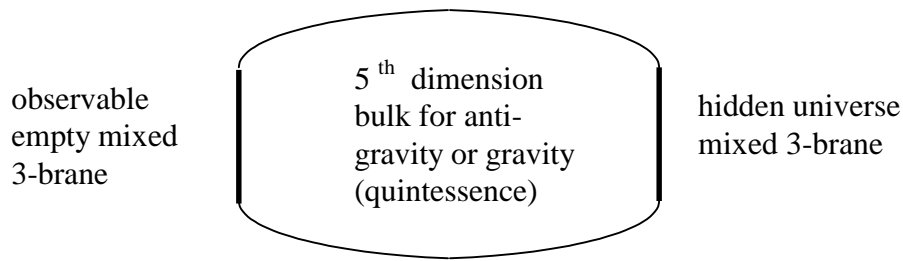

**Fig. 6:** Quintessence

It is anti-gravity because the franctionalization of the hidden universe is expansion. This anti-gravity is massless quintessence, causing the late cosmic accelerating expansion in the observable universe.

After a certain period, the hidden universe starts the condensation (contraction) phase. The hidden universe mixed 3-brane starts to condense into mixed 4-brane, inducing gravity in the fifth dimension bulk space. Consequently, quintessence in the observable universe causes the cosmic accelerating contraction. The quintessence transition involves both the accelerating expansion and the accelerating contraction for the observable universe. Quintessence controls the rate of expansion and contraction for the observable universe during the quintessence period to make them contracting at the same rate, so eventually both the observable universe and the hidden universe end at the same time.

When all hidden universe mixed 3-branes are converted into mixed 4-branes, the bulk space ceases to exist, and the observable universe starts to contract in a normal rate. At the end of the contraction, the big crush transition occurs.

The observable universe and the hidden universe form the boundary observable black hole and the boundary hidden 9-mixed brane separated by a finite gap (bulk space) for gravity. As gravity transfers from the boundary mixed branes to the bulk space, the mixed branes start to migrate to the bulk space for the interbrane de-mixing.

During this interbrane de-mixing, mixed branes are reversed back to unmixed branes, and the adjacent vacuum is reversed back to the bulk space, forming the pre-inflationary universe with a new set of space-time, separated from the space-time of the observable-hidden inflationary universe, as Fig. 7.

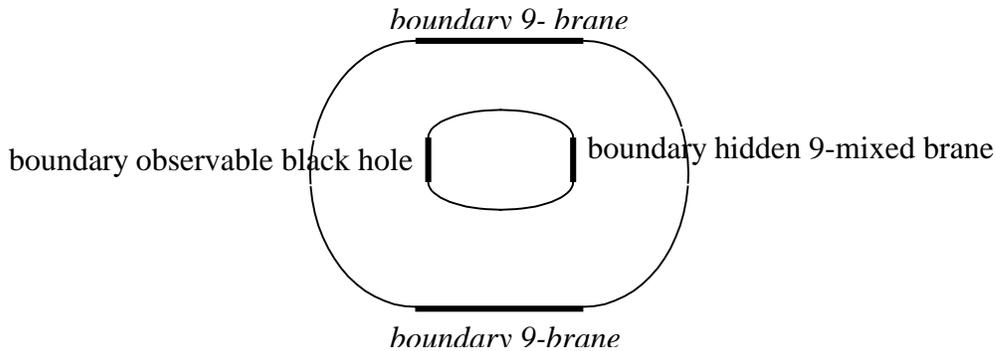

**Fig. 7:** the two boundary mixed branes in the inflationary universe and the two boundary 9-branes in the pre-inflationary universe

This pre-inflationary can then start another cycle of the universe. The cyclic universe is goes through the six transitions: the pre-inflation, the mixing, the big bang-band, the quintessence, the big crush, and the de-mixing as Fig. 8. The universe is an endless fattening free lunch.

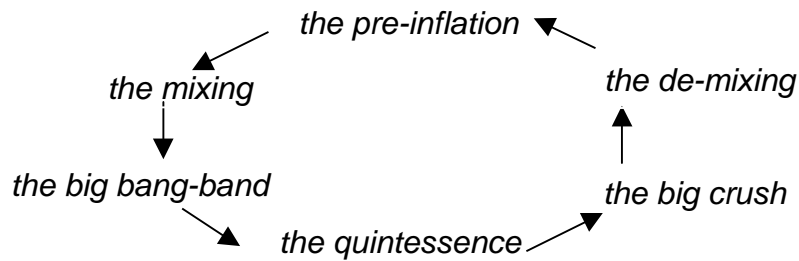

**Fig. 8**: cyclic universe: the six cosmic transitions

## 5. *Conclusion*

In this universe, there are three different types of vacuum: the adjacent vacuum, the adjacent zero space, and the empty space. The pre-universe vacuum brings about the pre-inflationary universe, containing the bulk space, which is the pre-adjacent vacuum. The absorption of the pre-adjacent vacuum by the pre-matter brings about the inflationary universe. The pre-adjacent vacuum emitted in the inflationary universe is the adjacent vacuum as the added space for the inflation in the inflationary universe, whose space-time is separated from the pre-inflationary universe. The empty space is generated as the gap among the mixed branes during the big bang inflation. The absorption of the adjacent vacuum as the added space results in the empty space superimposed with the adjacent zero space (no space), confining the matter in the form of particle. When the absorbed vacuum is emitted, it can be anywhere instantly in the empty space superimposed with the adjacent zero space where any point can be the starting point (zero point) of space-time. Consequently, the matter that expands into the adjacent vacuum has the probability to be anywhere instantly in the form of wavefunction. These three types of vacuum are included in the vacuum universe model, from which quantum mechanics and the periodic table of elementary particles are constructed to account for all elementary particles and hadrons and their masses in a good agreement with the observed values.